\documentclass[runningheads]{llncs}
\usepackage[T1]{fontenc}
\usepackage{graphicx}
\begin{document}
\title{Bias and Reliability in Learning-Style Inference from Student--AI Dialogues:
An Evaluation of GPT-4o and Gemini 2.0 Flash}
\titlerunning{Bias and Reliability in Learning-Style Inference}
\author{
Samaneh Zamanifard\inst{1} \and
Sajad Goudarzi\inst{1} \and
Andrew Robb\inst{1}
}

\authorrunning{Samaneh Zamanifard et al.}

\institute{
Clemson University, Clemson, USA.\\
\email{szamani@clemson.edu, sgoudar@clemson.edu, arobb@clemson.edu}
}

\maketitle              % typeset the header of the contribution

\begin{abstract}
Large Language Models (LLMs) are increasingly integrated into educational technologies for personalized support, creating opportunities for real-time learner modeling from dialogue. This study evaluates whether GPT-4o and Gemini 2.0 Flash can infer three Felder--Silverman Learning Style Model dimensions---Active--Reflective, Sensing--Intuitive, and Sequential--Global---from short, task-focused student--AI problem-solving dialogues. Using questionnaire-derived Index of Learning Styles labels from 153 undergraduate participants, we compared deterministic zero-shot and few-shot prompting under five-fold cross-validation. Although accuracy and support-weighted F1 were moderate for Sensing–Intuitive and Sequential–Global, majority-class baselines, macro-F1, balanced accuracy, and per-class recall showed that performance was often driven by dominant-label prediction rather than reliable recognition of both classes. The models substantially under-identified Intuitive and Global learners; few-shot prompting produced mixed effects and sometimes amplified majority-class bias. Active--Reflective classification remained unstable across models. These findings indicate that LLM-generated learner-style labels should be treated as uncertain, bias-prone signals rather than stable learner profiles, and that dialogue-based personalization systems require bias-aware evaluation and uncertainty-sensitive design.

\keywords{Artificial intelligence in education \and large language models \and learner modeling \and learning analytics \and fairness}
\end{abstract}
\section{Introduction}
Personalized learning aims to adapt instructional content, feedback, and support to learners' needs, behaviors, and preferences \cite{pane2017does}. AI-driven adaptive systems have advanced this goal by using learner models to guide content selection, assessment, and feedback \cite{pardamean2022ai}. Yet many existing approaches still depend on explicit questionnaires, diagnostic tests, or structured traces such as clickstream data \cite{hasibuan2022detection}. These sources are useful, but they can be burdensome, delayed, or incomplete in authentic learning environments, where students often reveal their reasoning through natural dialogue. Recent large language models (LLMs) create new opportunities for conversational learner modeling. Unlike conventional adaptive systems, LLM tutors can engage students in open-ended problem solving, respond to their questions, and generate feedback from the dialogue itself \cite{kasneci2023chatgpt,park2024empowering}. However, it remains unclear whether LLMs can reliably infer learner-related attributes from short educational conversations or whether their predictions reflect surface-level cues, class imbalance, or model-specific biases. This uncertainty is important because incorrect learner classifications may lead adaptive systems to provide misaligned feedback or support.

We examine this issue using the Felder--Silverman Learning Style Model (FSLSM), which represents learner preferences along Active--Reflective, Sensing--Intuitive, Visual--Verbal, and Sequential--Global dimensions \cite{felder1988learning,soloman2005index}. FSLSM has been widely used in adaptive learning research \cite{graf2006representative}. In this study, we use its dimensions as structured analytical labels for evaluating LLM behavior, not as claims that learning styles are fixed cognitive traits. Because our data consist only of text-based student--AI dialogues, we exclude the Visual--Verbal dimension and focus on Active--Reflective, Sensing--Intuitive, and Sequential--Global. This study evaluates GPT-4o and Gemini~2.0~Flash on 153 undergraduate student--AI problem-solving dialogues. Using Index of Learning Styles questionnaire-derived reference labels, we compare zero-shot and few-shot prompting and examine both aggregate performance and class-level misclassification patterns. The study contributes an empirical evaluation of dialogue-based learner-style inference, a bias-aware analysis of model behavior across prompting strategies, and evidence on the limitations of using LLM outputs as learner-modeling signals in adaptive educational systems. Accordingly, we address two research questions:
\begin{itemize}
\item \textbf{RQ1:} How accurately can GPT-4o and Gemini~2.0~Flash classify students' FSLSM dimensions from text-based problem-solving dialogues in a zero-shot setting?
\item \textbf{RQ2:} Does few-shot prompting improve classification performance and reduce bias compared with zero-shot prompting?
\end{itemize}
By analyzing both performance and systematic bias, this study informs the design of LLM-based personalization systems that treat inferred learner profiles as uncertain signals rather than definitive learner traits.

\section{Method}
\subsection{Chatbot Design and Implementation}
For this study, we developed a custom AI-powered chatbot using the GPT-4o-mini API. The chatbot was used only to collect student problem-solving dialogues; the later classification experiments with GPT-4o and Gemini~2.0~Flash were conducted offline on the saved dialogue data. The system was designed to support structured problem-solving dialogues and to capture participant interactions for later analysis. Because problem-solving tasks can promote engagement, reasoning, and learning in online environments \cite{hendarwati2021collaborative}, participants selected a topic from domains such as Artificial Intelligence, Cybersecurity, Data Science, Environmental Science, Game Development, and Psychology. For each topic, the chatbot presented an open-ended scenario involving a central concept, multiple contributing factors, and several possible solution paths.

To encourage substantive interaction, the interface disabled copy-and-paste functionality, and the chatbot prompted for clarification when participants attempted to submit a final answer without prior dialogue. Participants were required to complete at least ten acceptable dialogue turns. An acceptable turn was defined as a meaningful contribution to the problem-solving exchange, including proposing an idea, explaining reasoning, responding to a prompt, or addressing a follow-up question. Off-topic messages, direct requests for final answers, and attempts to manipulate the chatbot were recorded but excluded from the minimum-turn count and from the analytical dataset during preprocessing.

\subsection{Participants}

A total of 270 undergraduate students from a large public university in the southeastern United States participated in the study. Eligible participants were at least 18 years old, enrolled as undergraduates, and had prior experience using LLM-based tools for learning. Students were recruited through in-class announcements and received extra credit, with an equivalent alternative assignment available for those who did not participate. Participation was voluntary, and all procedures were approved by the university's Institutional Review Board (IRB).

\subsection{Data Collection}

The experiment included three phases. First, participants reviewed the IRB-approved consent form and confirmed their understanding of the study procedures. Second, they completed the 44-item ILS questionnaire \cite{soloman2005index}, which was used as the reference-label source for ILS-based learning-preference labels. For each ILS dimension, the higher-scoring side was assigned as the participant's label. To verify label integrity, two researchers independently audited a random 15\% subset of records and confirmed that questionnaire responses were correctly matched to the assigned labels. In the final phase, participants selected a topic of interest and completed a problem-solving dialogue with the chatbot. Across the original 270 participants, the system recorded 2,873 student message turns, with an average of 10.64 turns per participant. Because the interaction was entirely text-based, the Visual--Verbal dimension was excluded, leaving Active--Reflective, Sensing--Intuitive, and Sequential--Global for model evaluation. Table~\ref{tab:learning_style_summary} summarizes the resulting label distribution. Reflecting patterns commonly reported in FSLSM-based studies of student populations \cite{soloman2005index,brown2009learning}, our sample was skewed toward Sensing and Sequential preferences, while the Active--Reflective dimension was nearly balanced. To ensure data quality, two authors manually reviewed all interaction logs using predefined exclusion criteria. Dialogues were excluded if they contained fewer than ten acceptable student turns, incomplete or unmatched ILS responses, technical logging failures or duplicate records, predominantly off-task or superficial exchanges, direct answer-seeking without substantive reasoning, or attempts to manipulate the chatbot. Disagreements were resolved through discussion. Of the 270 original participants, 153 were retained for analysis, and 117 were excluded for meeting one or more predefined exclusion criteria: fewer than ten acceptable turns, incomplete or unmatched ILS records, technical logging failures or duplicate records, predominantly off-task or superficial exchanges, direct answer-seeking without substantive reasoning, or attempts to manipulate the chatbot. The final analytical dataset contained 153 participants and 1,650 student message turns.

\begin{table}
\caption{ILS-based reference-label distribution and majority-class baselines.}
\label{tab:learning_style_summary}
\centering
\begin{tabular}{|l|l|l|l|}
\hline
\textbf{Dimension} & \textbf{Class A} & \textbf{Class B} & \textbf{Majority baseline} \\
\hline
ACT--REF & ACT: 80 (52.3\%) & REF: 73 (47.7\%) & 52.3\% \\
SEN--INT & SEN: 114 (74.5\%) & INT: 39 (25.5\%) & 74.5\% \\
SEQ--GLO & SEQ: 98 (64.1\%) & GLO: 55 (35.9\%) & 64.1\% \\
\hline
\end{tabular}

\vspace{2mm}
\footnotesize{\textit{Note.} The majority baseline is the accuracy obtained by always predicting the more frequent class for each dimension.}
\end{table}

\subsection{Evaluation Metrics}

We evaluated zero-shot and few-shot classification using accuracy, support-weighted F1, macro-F1, balanced accuracy, and per-class recall. Because SEN–INT and SEQ–GLO were imbalanced, accuracy and support-weighted averages were interpreted relative to the empirical majority-class baseline for each dimension. Macro-F1 and balanced accuracy were treated as the primary reliability metrics because they give equal weight to both labels and reveal whether performance reflects recognition of both classes rather than dominant-label prediction. Weighted recall is not reported because, in single-label classification, it is identical to accuracy. For each dialogue, the model generated predictions for all three retained FSLSM dimensions simultaneously: Active–Reflective, Sensing–Intuitive, and Sequential–Global. We used participant-level five-fold cross-validation \cite{geisser1975predictive,hastie2009elements}. In each iteration, one fold served as the validation set, while the remaining four folds were used only as the example pool for the few-shot condition. The zero-shot condition used the same validation folds without labeled examples, allowing direct comparison between prompting strategies. Final metrics were averaged across the five folds. We computed 95\% bootstrap confidence intervals for support-weighted F1 using 1,000 participant-level resamples with random seed 42. The confidence intervals, therefore, quantify uncertainty in weighted F1, not in accuracy, macro-F1, or balanced accuracy.

\subsection{Prompt Engineering}
Prompt design can substantially affect LLM performance in reasoning and classification tasks \cite{brown2020language,chae2026large}. Therefore, we iteratively refined the prompts for both zero-shot and few-shot settings to improve consistency across model outputs. All experiments used deterministic decoding (temperature = 0) to reduce stochastic variation and support reproducible classification \cite{gilardi2023chatgpt}. Prompt refinement was conducted on a small pilot subset of 12 randomly selected samples, with representation across learning-style categories. At each iteration, we manually compared model predictions with the questionnaire-derived ILS labels and examined common errors, including confusion between opposing categories and logically unsupported classifications. Initial prompts asked the models to infer numerical ILS-style scores for each opposing pair, with each pair summing to 11. However, these score-based predictions were unstable and often inconsistent with the reference labels. We therefore reformulated the task as categorical classification, requiring the model to assign one label from each retained dimension: Active or Reflective, Sensing or Intuitive, and Sequential or Global. This categorical framing produced more consistent outputs during manual review and is consistent with recent findings that LLMs perform more reliably on categorical classification than on continuous numerical prediction tasks \cite{avnat2025performance}.

\section{Results}
This section evaluates the zero-shot and few-shot classification performance of GPT-4o and Gemini 2.0 Flash in inferring learners' ILS-based preferences from their conversational interactions with the chatbot. Model performance was assessed using accuracy, support-weighted F1, macro-F1, balanced accuracy, per-class recall, and 95\% bootstrap confidence intervals for support-weighted F1. Because the SEN--INT and SEQ--GLO dimensions were imbalanced, accuracy was interpreted relative to the empirical majority-class baselines in Table~\ref{tab:learning_style_summary}, rather than against a nominal 50\% chance level. The 95\% confidence intervals reported in Table~\ref{tab:combined_zero_few_performance} refer to support-weighted F1.

\subsection{Zero-shot Performance}
In the zero-shot setting, learning styles were predicted using only pre-trained model knowledge, without task-specific examples. This subsection focuses on the zero-shot rows in Table~\ref{tab:combined_zero_few_performance}.

On SEN--INT, both models produced their highest aggregate zero-shot performance, but neither exceeded the 74.5\% majority-class baseline. GPT-4o achieved 70\% accuracy and 68\% support-weighted F1, while Gemini 2.0 Flash achieved 73\% accuracy and 68\% support-weighted F1. Macro-F1 and balanced accuracy were substantially lower than weighted F1, indicating uneven recognition of the two classes.

On SEQ--GLO, GPT-4o slightly exceeded the 64.1\% majority-class baseline, achieving 67\% accuracy and 62\% support-weighted F1. Gemini 2.0 Flash approximately matched the majority baseline, with 64\% accuracy and 56\% support-weighted F1. Macro-F1 and balanced accuracy remained low, especially for Gemini, suggesting that aggregate scores masked weak balanced classification.

On ACT--REF, both models showed limited reliability. GPT-4o achieved 58\% accuracy and 54\% support-weighted F1, only modestly above the 52.3\% majority-class baseline. Gemini 2.0 Flash achieved 53\% accuracy and 40\% support-weighted F1, approximately matching the majority baseline in accuracy while showing poor macro-F1. These results suggest that conversational signals distinguishing Active from Reflective tendencies were not consistently captured in the zero-shot setting.

% Add this in the preamble if it is not already included:
% \usepackage{graphicx}

\begin{table}
\caption{Aggregate and balanced performance of LLMs across zero-shot and few-shot settings at zero temperature.}
\label{tab:combined_zero_few_performance}
\centering
\scriptsize
\begin{tabular}{|l|l|l|l|l|l|l|l|l|l|}
\hline
\textbf{Setting} & \textbf{Model} & \textbf{Dimension} &
\textbf{ACC} & \textbf{Maj.} & \textbf{F1} &
\textbf{Macro-F1} & \textbf{B-Acc} &
\textbf{95\% CI} \\
\hline

Zero-shot & GPT-4o & ACT--REF & 58\% & 52.3\% & 54\% & 54\% & 59\% & [45\%, 62\%] \\
Zero-shot & GPT-4o & SEN--INT & 70\% & 74.5\% & 68\% & 57\% & 56\% & [60\%, 77\%] \\
Zero-shot & GPT-4o & SEQ--GLO & 67\% & 64.1\% & 62\% & 56\% & 58\% & [54\%, 71\%] \\
\hline

Zero-shot & Gemini 2.0 Flash & ACT--REF & 53\% & 52.3\% & 40\% & 38\% & 51\% & [30\%, 49\%] \\
Zero-shot & Gemini 2.0 Flash & SEN--INT & 73\% & 74.5\% & 68\% & 52\% & 53\% & [58\%, 76\%] \\
Zero-shot & Gemini 2.0 Flash & SEQ--GLO & 64\% & 64.1\% & 56\% & 47\% & 52\% & [47\%, 64\%] \\
\hline

Few-shot & GPT-4o & ACT--REF & 56\% & 52.3\% & 55\% & 54\% & 55\% & [47\%, 62\%] \\
Few-shot & GPT-4o & SEN--INT & 73\% & 74.5\% & 63\% & 42\% & 49\% & [53\%, 72\%] \\
Few-shot & GPT-4o & SEQ--GLO & 57\% & 64.1\% & 50\% & 41\% & 46\% & [40\%, 59\%] \\
\hline

Few-shot & Gemini 2.0 Flash & ACT--REF & 54\% & 52.3\% & 45\% & 44\% & 52\% & [36\%, 54\%] \\
Few-shot & Gemini 2.0 Flash & SEN--INT & 73\% & 74.5\% & 67\% & 50\% & 52\% & [57\%, 75\%] \\
Few-shot & Gemini 2.0 Flash & SEQ--GLO & 61\% & 64.1\% & 58\% & 51\% & 53\% & [50\%, 67\%] \\
\hline

\end{tabular}

\vspace{2mm}
\footnotesize{\textit{Note.} Maj. denotes the empirical majority-class accuracy baseline for each learning-style dimension. F1 is the support-weighted F1-score. Macro-F1 is the unweighted average of the two classwise F1-scores. B-Acc denotes balanced accuracy. The 95\% confidence interval (CI) is reported for the weighted F1-score.}
\end{table}

\subsection{Zero-shot Bias Patterns from Confusion Matrices}

The zero-shot confusion matrices show that aggregate performance was often driven by class-preference effects rather than balanced recognition of both labels. On SEN--INT, both models achieved their best aggregate scores yet favored SEN predictions: GPT-4o correctly identified 96/114 SEN learners (84.2\%) but only 11/39 INT learners (28.2\%), while Gemini 2.0 Flash identified 107/114 SEN learners (93.9\%) but only 5/39 INT learners (12.8\%). On SEQ--GLO, both models again favored SEQ predictions: GPT-4o identified 90/98 SEQ learners (91.8\%) but only 13/55 GLO learners (23.6\%), while Gemini identified 92/98 SEQ learners (93.9\%) but only 6/55 GLO learners (10.9\%). On ACT--REF, the bias direction differed across models. GPT-4o favored REF learners, identifying 65/73 REF learners (89.0\%) but only 23/80 ACT learners (28.8\%). In contrast, Gemini favored ACT learners, identifying 78/80 ACT learners (97.5\%) but only 3/73 REF learners (4.1\%). These patterns show that zero-shot predictions were strongly affected by model-specific and class-specific biases.

\subsection{Few-shot Performance in Learning-Style Classification}
To evaluate whether few-shot prompting improves classification, the models were provided with labeled examples for each learning style category. We used 5-fold cross-validation \cite{geisser1975predictive,hastie2009elements}, where in each iteration one fold served as the validation set and the remaining four folds supplied few-shot examples. Due to LLM context-window constraints, 60 examples were randomly sampled from the training folds for each iteration. Although the 60 exemplars were selected at random, they were checked to ensure that each learning-style category was represented as evenly as possible, given the underlying class distribution.  Performance metrics were averaged across folds and compared to zero-shot results. The few-shot rows in Table~\ref{tab:combined_zero_few_performance} summarize aggregate and balanced performance metrics.

% \begin{table}
% \caption{Few-shot weighted performance of LLMs at zero temperature.}
% \label{tab:fewshot_performance_zero_temperature}
% \centering
% \begin{tabular}{|l|l|l|l|l|l|l|}
% \hline
% \textbf{Model} & \textbf{Learning Style} & \textbf{ACC} & \textbf{PRE} & \textbf{REC} & \textbf{F1} & \textbf{95\% CI} \\
% \hline
% GPT-4o & ACT-REF & 56\% & 55\% & 56\% & 55\% & [47\%, 62\%] \\
% GPT-4o & SEN-INT & 73\% & 55\% & 73\% & 63\% & [53\%, 72\%] \\
% GPT-4o & SEQ-GLO & 57\% & 47\% & 57\% & 50\% & [40\%, 59\%] \\
% \hline
% Gemini 2.0 Flash & ACT-REF & 54\% & 55\% & 54\% & 45\% & [36\%, 54\%] \\
% Gemini 2.0 Flash & SEN-INT & 73\% & 67\% & 73\% & 67\% & [57\%, 75\%] \\
% Gemini 2.0 Flash & SEQ-GLO & 61\% & 57\% & 61\% & 58\% & [50\%, 67\%] \\
% \hline
% \end{tabular}
% \end{table}

On SEN--INT, few-shot prompting did not improve balanced performance. GPT-4o accuracy increased from 70\% to 73\%, but this remained below the 74.5\% majority-class baseline, and support-weighted F1 decreased from 68\% to 63\%. Gemini 2.0 Flash remained close to its zero-shot aggregate performance, with 73\% accuracy and 67\% support-weighted F1, but it also remained below the majority-class baseline. For both models, macro-F1 and balanced accuracy were substantially lower than support-weighted F1, indicating that few-shot prompting did not improve reliable recognition of both SEN and INT learners.

On SEQ--GLO, GPT-4o's performance declined under few-shot prompting. Accuracy decreased from 67\% to 57\%, and support-weighted F1 dropped from 62\% to 50\%, with the F1 confidence interval shifting from [54\%, 71\%] to [40\%, 59\%]. Gemini 2.0 Flash showed a small increase in support-weighted F1 from 56\% to 58\%, with the confidence interval shifting from [47\%, 64\%] to [50\%, 67\%]. However, Gemini's accuracy remained below the 64.1\% majority-class baseline, and balanced accuracy remained low. Therefore, the Gemini change should be interpreted as a small aggregate shift rather than a stable improvement in balanced SEQ--GLO classification.

Few-shot prompting produced limited changes on ACT--REF. GPT-4o's support-weighted F1 increased only from 54\% to 55\%, with accuracy changing from 58\% to 56\%. Gemini 2.0 Flash increased from 40\% to 45\% support-weighted F1, with accuracy changing from 53\% to 54\%. However, macro-F1 and balanced accuracy remained relatively low, indicating that few-shot prompting did not substantially improve classification reliability for this dimension.

Overall, few-shot prompting did not consistently improve performance across dimensions. Improvements were minimal for Gemini and sometimes detrimental for GPT-4o, particularly on SEQ--GLO. These findings suggest that adding labeled examples does not guarantee stronger or more reliable inference of learning styles from short educational dialogues.

\subsection{Few-shot Bias Patterns from Confusion Matrices}
The few-shot confusion matrices show that few-shot prompting frequently shifted, and in some cases intensified, the class-preference biases observed in the zero-shot condition. On SEN--INT, both models further increased their bias toward Sensing learners. GPT-4o classified 112/114 SEN learners correctly (98.2\%) but failed to identify any INT learners (0/39). Gemini similarly identified 108/114 SEN learners (94.7\%) but only 4/39 INT learners (10.3\%). These results show that few-shot prompting amplified majority-class dependence rather than correcting it.

On SEQ--GLO, GPT-4o's recall decreased for both groups, correctly identifying 83/98 SEQ learners (84.7\%) and only 4/55 GLO learners (7.3\%). Gemini showed modest improvement in recognizing GLO learners, identifying 12/55 GLO learners (21.8\%), while still identifying SEQ learners at a much higher rate, 82/98 (83.7\%). Despite this improvement, both models continued to favor SEQ over GLO. On ACT--REF, GPT-4o reversed its zero-shot bias: few-shot prompting shifted its preference toward ACT learners, with 55/80 ACT learners identified (68.8\%) compared with 30/73 REF learners (41.1\%). Gemini retained its tendency to favor ACT predictions, although this tendency weakened slightly: ACT recall decreased to 73/80 (91.3\%), while REF recall improved to 9/73 (12.3\%). Taken together, the few-shot confusion matrices show that providing labeled examples did not reliably reduce bias. Instead, it often intensified dominant-class preferences, reversed bias direction, or increased classification instability.

\section{Discussion}

LLMs are increasingly used in educational technologies for personalized feedback, content delivery, and interactive tutoring \cite{kasneci2023chatgpt,park2024empowering,baidoo2023education}. Yet learner modeling in many adaptive systems still relies on explicit self-report instruments such as the ILS, which can be difficult to administer repeatedly or at scale. This study examined whether GPT-4o and Gemini~2.0~Flash can infer ILS-based learning-style tendencies from short, task-focused dialogues without explicit questionnaires (RQ1), and whether few-shot prompting improves this inference compared with zero-shot prompting (RQ2). Because we use ILS labels as structured analytical categories rather than fixed cognitive traits, our focus is not only classification performance, but also reliability, class imbalance effects, and systematic bias.

\subsection{Model Behavior Under Zero-Shot and Few-Shot Prompting}

The zero-shot results suggest that the models may have captured limited dialogue-associated cues, but the stronger finding is systematic class preference. On SEN–INT and SEQ–GLO, apparently moderate weighted scores were largely explained by overprediction of Sensing and Sequential labels. In several conditions, accuracy was at or below the empirical majority-class baseline, and balanced accuracy remained weak. Therefore, these outputs should not be interpreted as reliable learning-style detection. ACT–REF was even less stable: GPT-4o and Gemini 2.0 Flash showed opposite bias directions, indicating that short text-based dialogues did not provide robust cues for this dimension.

These biases may reflect an interaction between the sample distribution and broader model-level priors. Our dataset was skewed toward Sensing and Sequential learners, and prior FSLSM studies report similar patterns in technical student populations \cite{soloman2005index,brown2009learning}. Because LLMs are pretrained on large-scale web and textual corpora that likely include educational materials, tutorials, worked examples, programming resources, and prior learning-style research, they may encode dominant patterns in which concrete and stepwise reasoning appears more typical. More generally, LLMs can reproduce or amplify distributional biases present in their training data \cite{gallegos2024bias}. Therefore, the overprediction of Sensing and Sequential labels should not be interpreted only as successful learner-style detection; it may also reflect learned distributional bias.

Few-shot prompting did not consistently resolve these issues. Although labeled examples can help LLMs attend to task-relevant cues, their effectiveness is highly task- and context-dependent \cite{alizadeh2025open}. In our results, few-shot prompting produced mixed effects: some metrics shifted modestly, but majority-class tendencies often persisted or intensified. For GPT-4o, few-shot prompting reduced performance on Sequential--Global and strengthened overprediction patterns on Sensing--Intuitive. For Gemini~2.0~Flash, gains were limited and accompanied by continued class imbalance effects. On Active--Reflective, few-shot prompting produced only marginal improvement and did not yield stable classification behavior. Overall, few-shot prompting did not reliably reduce bias; instead, it introduced a trade-off between small metric changes and more pronounced or unpredictable class preferences.

Taken together, the findings answer RQ1 and RQ2 by showing that off-the-shelf LLMs can recover limited dialogue-associated signals, but not robust learner-style classifications. The strongest aggregate scores were largely explained by dominant-label prediction, especially for Sensing and Sequential labels. Few-shot prompting did not reliably reduce this bias and sometimes intensified it. Therefore, zero-shot and few-shot outputs should be treated as tentative, bias-prone indicators rather than stable learner profiles.

\subsection{Implications for Fair and Responsible LLM-Based Personalization}

The recurring bias toward majority classes raises concerns for educational deployment. If an adaptive system misidentifies a learner's preference, it may provide misaligned content, feedback, or support \cite{boateng2025algorithmic,vaida2020brief}. For example, a Global learner misclassified as Sequential may receive overly linear guidance when they would benefit from seeing the broader structure of a problem. Such risks align with broader concerns about fairness \cite{gao2024fairness,tiwari2025biases}, accountability \cite{quttainah2024cost}, and transparency \cite{mienye2025chatgpt,singh2025evaluating} in educational AI.

Some misclassification risk also comes from categorical operationalization. The ILS produces continuous scores, but this study, like much work using ILS in adaptive learning systems \cite{graf2006representative,graf2007depth}, converts those scores into binary labels. This makes the task suitable for classification, but it can exaggerate small differences near the cutpoint and make borderline learners difficult to classify. Future systems should therefore consider confidence scores, soft labels, or ordinal representations that better reflect uncertainty in learner profiles.

Responsible use of LLM-based learner modeling requires safeguards. Educators and learners should be able to review, question, or override model-generated classifications, ideally alongside complementary evidence such as survey responses, performance data, or interaction history. Personalization should also avoid rigid routing based on a single predicted style. Instead, systems should offer multiple representations of content, such as both step-by-step explanations and high-level overviews, so that misclassification does not restrict learners' options. Finally, educational AI systems should communicate uncertainty explicitly, ensuring that inferred profiles support learner agency rather than replacing it.

\subsection{Limitations}

This study has several limitations. First, ILS labels were derived from self-report questionnaire responses and should be interpreted as reference labels rather than objective learner traits. Second, the analytical dataset was limited to 153 undergraduate participants from a single institutional context, and SEN–INT and SEQ–GLO were imbalanced. Third, the study used short text-based dialogues and excluded the Visual–Verbal FSLSM dimension. Fourth, prompt refinement, few-shot exemplar selection, output parsing, and provider-side model updates may affect reproducibility. Fifth, the study did not compare LLM prompting against supervised non-LLM text classifiers, so the findings should not be interpreted as showing that LLMs are preferable to simpler learner-modeling baselines. Finally, results may differ with other models, repeated exemplar sampling, calibrated classifiers, soft-label representations, longer dialogues, or hybrid learner-modeling approaches.

\section{Conclusion}

Using ILS questionnaire-derived reference labels, this study examined whether GPT-4o and Gemini~2.0~Flash can infer students' learning-preference indicators from short educational chatbot dialogues. For RQ1, the models showed moderate accuracy and support-weighted F1 on SEN–INT and SEQ–GLO, but these scores were often at or below majority-class baselines and masked low minority-class recall. Macro-F1, balanced accuracy, and per-class recall showed that performance was frequently driven by dominant-label prediction rather than reliable recognition of both classes. Active–Reflective classification remained unstable and model-dependent. For RQ2, few-shot prompting produced mixed effects and did not reliably reduce bias; in several cases, it intensified majority-class preferences or degraded balanced performance. Overall, LLM-generated learner-style classifications should be treated as provisional, uncertain signals rather than stable learner traits. Dialogue-based learner modeling may capture context-dependent communication patterns, but responsible use in adaptive learning systems requires uncertainty-aware design, transparency, learner control, and safeguards against the under-identification of minority profiles.

\bibliographystyle{splncs04}
\bibliography{bibliography}


\end{document}